\documentclass[12pt,preprint,longabstract]{aastex}

\shorttitle{Radio Loud and Radio Quiet QSOs}
\shortauthors{Kellermann et al.}

\begin{document}

\title{Radio Loud and Radio Quiet QSOs}

\author{K.~I.~Kellermann\altaffilmark{1} and J.~J.~Condon\altaffilmark{1}}
\affil{National Radio Astronomy Observatory, 520 Edgemont Rd., 
Charlottesville, VA 22903, USA}
\email{kkellerm@nrao.edu}

\author{A.~E.~Kimball\altaffilmark{1} and R.~A.~Perley\altaffilmark{1}}
\affil{National Radio Astronomy Observatory, Socorro, NM 87801, USA}

\and

\author{\v{Z}eljko Ivezi\'{c}}
\affil{Department of Astronomy, University of Washington, Box 351580,
  Seattle, WA 98195, USA}

\altaffiltext{1}{The National Radio Astronomy Observatory is a
  facility of the National Science Foundation operated under
  cooperative agreement by Associated Universities, Inc.}

\begin{abstract}

We discuss 6 GHz JVLA observations covering a volume-limited sample of
178 low redshift ($0.2 < z < 0.3$) optically selected QSOs.  Our 176
radio detections fall into two clear categories: (1) About $20$\% are
radio-loud QSOs (RLQs) having spectral luminosities $L_6 \gtrsim
10^{\,23.2} \mathrm{~W~Hz}^{-1}$ primarily generated in the active
galactic nucleus (AGN) responsible for the excess optical luminosity
that defines a \emph{bona fide} QSO. (2) The radio-quiet QSOs (RQQs)
have $10^{\,21} \lesssim L_6 \lesssim 10^{\,23.2} \mathrm{~W~Hz}^{-1}$
and radio sizes $\lesssim 10 \mathrm{~kpc}$, and we suggest that the
bulk of their radio emission is powered by star formation in their
host galaxies.  ``Radio silent'' QSOs ($L_6 \lesssim 10^{\,21}
\mathrm{~W~Hz}^{-1}$) are rare, so most RQQ host galaxies 
form stars faster than the Milky Way; they are not ``red
and dead'' ellipticals.  Earlier radio observations did not have the
luminosity sensitivity $L_6 \lesssim 10^{\,21} \mathrm{~W~Hz}^{-1}$
needed to distinguish between such RLQs and RQQs. Strong, generally double-sided,
radio emission spanning $\gg 10 \mathrm{~kpc}$ was found
associated with 13 of the 18 RLQ cores having peak flux densities
$S_\mathrm{p} > 5 \mathrm{~mJy~beam}^{-1}$ ($log(L) \gtrsim 24$).  The radio luminosity
function of optically selected QSOs and the extended radio emission
associated with RLQs are both inconsistent with simple ``unified''
models that invoke relativistic beaming from randomly oriented QSOs to
explain the difference between RLQs and RQQs.  Some intrinsic property
of the AGNs or their host galaxies must also determine whether or not
a QSO appears radio loud.
\end{abstract}


\section{Introduction}

Fifty years after the discovery of optically selected QSOs, the
distinction between radio-loud QSOs (RLQs) and radio-quiet QSOs (RQQs)
is still being debated.   In a preliminary report, \citet{KKC11} discussed 6 GHz observations of an optically selected QSO sample taken from the SDSS. These observations, made with the Karl G. Jansky Very Large Array (VLA), achieved unprecedented sensitivity which led to the detection of 97\% of the  SDSS QSOs
with $M_\mathrm{i} < -23$ in a low-redshift ($0.2 < z < 0.3$)
volume-limited sample.  In this paper, we report the full results of these observations including a new data analysis that detects
99\% of the 178 studied  QSOs
with $M_\mathrm{i} < -23$ in a low-redshift ($0.2 < z < 0.3$)
volume-limited sample. Section~\ref{sec:background} reviews the
different criteria used to distinguish RLQs from RQQs and the evidence
that RLQs and RQQs are or are not distinct populations.  In
Section~\ref{sec:JVLAsample} we describe our volume-limited sample of
low-redshift QSOs and present results from the VLA observations that
suggests that for most
 QSOs, the low luminosity ($10^{21} \lesssim L_6 \lesssim 10^{23} \mathrm{~W~Hz}^{-1}$) radio emission is
primarily the result of star formation in the host galaxy, while 
the emission in the more luminous objects the radio emission is
related to the SMBH.  Section~\ref{sec:sourcenotes} includes notes on
individual sources.  In Section~\ref{sec:twopopulations} we discuss
differences between RLQs and RQQs.  Our conclusions are summarized in
Section~\ref{sec:summary}.

All calculations in this paper assume a flat $\Lambda$CDM cosmology
with $H_\mathrm{0}=70$ km s$^{-1}$ Mpc$^{-1}$ and
$\Omega_\Lambda=0.7$.  Spectral luminosities are specified by
their source-frame frequencies, flux densities are specified
in the observer's frame, and a mean spectral index $\alpha \equiv
d \log (S) / d \log (\nu) = -0.7$ is used to make frequency conversions 
(see \citet{KKC11} for details).

\section{Background}\label{sec:background}

Shortly following the recognition that quasi-stellar radio sources
were surprisingly luminous at both radio and optical wavelengths
\citep{S63,G63,GS64,SM65}, Allan \citet{S65} announced ``The Existence
of a Major New Constituent of the Universe: the Quasistellar
Galaxies'' which were optically selected and outnumbered the known
(and extremely radio loud) quasi-stellar radio sources by a factor of
$\sim 10^3$.  Sandage's dramatic announcement was published in the May
1, 1965 issue of the {\it Astrophysical Journal} (the same day that it
was received at the journal office) and created considerable
controversy. \citet{Z65} immediately pointed out that he had
previously reported on similar objects which he called ``blue compact
galaxies,'' while \citet{K65} and \citet{LV65} argued that most of
Sandage's claimed quasi-stellar galaxies are only blue stars in our
Galaxy.  Nevertheless, it is now widely recognized that Sandage's
quasi-stellar galaxies, today called quasi-stellar objects (QSOs), do
outnumber quasi-stellar radio sources, albeit only by about an order
of magnitude.

Although the term ``quasar'' is a portmanteau of ``quasi-stellar radio
source'' \citep{CHIU64}, here we follow the definitions given by
\citet{S70} and endorsed by the {\it Astrophysical Journal} that (1)
quasars include all extragalactic objects of ``starlike appearance (or
containing a dominant starlike component),'' even ``buried'' objects
reddened or obscured by dust and objects selected at infrared or X-ray
wavelengths, and (2) QSOs are quasars selected optically.
\citet{SG83} quantified the requirement for a ``dominant starlike
component'' to distinguish {\it bona fide} QSOs from less-luminous
AGNs such as Seyfert nuclei: QSOs must have $M_\mathrm{B} \leq -23$
when calculated using $H_0 = 50\,
\mathrm{km\,s}^{-1}\,\mathrm{Mpc}^{-1}$ and $q_0 = 0.1$.  In the
modern $\Lambda$CDM cosmology with $H_0 =
70\,\mathrm{km\,s}^{-1}\,\mathrm{Mpc}^{-1}$, $\Omega_\mathrm{m} =
0.3$, and $\Omega_\Lambda = 0.7$, such QSOs have absolute magnitudes
$M_\mathrm{B} \leq -22.2$.  However, most authors still use
$M_\mathrm{B} < -23$, following the letter but not the spirit of the
\citet{SG83} criterion, and we also use that $0.8 \mathrm{~mag}
\approx$ 2 times more luminous limit for QSOs.

Early targeted searches for radio emission from QSOs \citep{KPT66,
  SW66, K73,F77, SW80, SW80b, C80a, C80b} had $\sim 10$\% detection
rates.  Somewhat surprisingly, the radio detection rate grew very
slowly as the flux-density limit was lowered, leading to the
suggestion that radio-loud QSOs (RLQs) and radio-quiet QSOs (RQQs)
might have different flux-density distributions and thus comprise two
distinct populations.  (Note that radio quiet has always meant radio
weak, not completely radio silent).  \citet{SPSW80} discussed the
radio observations of 70 spectroscopically confirmed QSOs selected
from the Michigan Curtis Schmidt surveys \citep{MSL77a,MSL77b,MSL77c}
and claimed that the distribution of the logarithmic radio flux
densities $\log(S)$ of QSOs is sharply bimodal (that is, has two
peaks).  This stimulated a controversy that persists to this day.
However, \citet{CKK13} recently pointed out that this claim resulted
from a mathematical error and is not valid.

More sensitive 5 GHz VLA observations with an rms noise $\sigma
\approx 65 \, \mu \mathrm{Jy~beam}^{-1}$ allowed \citet{KSS89} to
detect 82\% of the relatively bright ($B \lesssim 16$) {\it bona fide}
QSOs with $M_\mathrm{B} < -23$ (old cosmology; $M_\mathrm{B} < -22.2$
today) in the \citet{SG83} Bright Quasar Survey (BQS).  They reached a
radio/optical flux-density ratio $R = 0.1$ and argued that the
distributions of observed $\log(S)$ and $\log(R)$ appeared to be
bimodal.  \citet{S70} preferred $R$ rather than $S$ because it is
nearly distance-independent.  Also, if the radio and optical
luminosites of QSOs are correlated, the distribution of $\log(R)$
should be more narrow than the distribution of $\log(S)$, making it a more
efficient discriminator between RLQs and RQQs.  The \citet{KSS89} VLA
observations of bright QSOs had sufficient sensitivity to divide most
of them into an RQQ population defined by $R < 10$ and an RLQ
population defined by $R > 10$, with few QSOs falling in the $1 < R <
10$ gap.  \citet{GMF92} and \citet{MRS93} noted that large errors in
the magnitudes, positions, and redshifts of BQS objects resulted in
substantial bias and incompleteness in the BQS, raising questions
about the conclusions of \citet{KSS89}.  Even so, several subsequent
radio observations and analyses of other QSO samples appear to support
their two-population description
\citep{MPM90,SMWF92,VIF92,GKMD99,XLB99,ILK02,IRH04,JFI07}.

However, the claim of two distinct populations was further challenged
by a number of other authors who found continuous distributions of
$\log(S)$ or $\log(R)$ with no clear gap separating RLQs from RQQs.
Using 1.4 GHz radio data from the FIRST survey, \citet{BWH95,
  WBG00, LLR01, CCC03, CMC03, WBC05, RCW09, SPLS11, BLM13} all reported no
evidence for two separate populations, although these conclusions were
all based on samples defined by radio surveys, which are highly biased
toward the radio-loud population.  Moreover, with a $1
\mathrm{\,mJy\,beam}^{-1}$ catalog limit, FIRST cannot detect RQQs
individually, so \citet{WHB07} lowered their noise limit by stacking
over 40,000 FIRST images of SDSS QSOs.  Unfortunately, the resulting
mean flux density is dominated by the very small number of sources
just below the FIRST catalog limit \citep{CKK13}. \citet{CCC03,CMC03}
argued that the apparent bimodality claimed by others is the result of
bias related to the strong dependence of radio on optical luminosity,
but this argument has been refuted by \citet{IRH04}.

Using their 20 GHz radio data, \citet{MSC12} found no evidence for a
separate population of RLQs and RQQs in a sample of 874 X-ray selected
quasars, but their X-ray selection biases the sample toward blazars and
the 20 GHz ATCA observations barely reach down into the RQQ
population.

Terlevich and colleagues \citep{TMM87,T92,TTFM92} were the first to
propose that nearly all QSO emission (radio, optical continuum, and
even the broad emission lines) is powered by starbursts rather than by
supermassive black holes (SMBHs).  For at least a few of their
quasars, \citet{KSS94,BBLG96,BB98,BLB05,UAB05}, and more recently \citet{RMN16} and \citet{MPN16}, presented
evidence for radio structures extending beyond the host galaxy of
stars, radio components with very high surface brightnesses,
variability, or apparent superluminal motion, all of which are
characteristic of radio emission driven by a SMBH.  However, given
that these quasars all lie near the \citet{KSS89} $R \sim 10$ boundary
between the RQQ and RLQ distributions, they do not provide evidence
that RQQs are SMBH-powered.

\citet{SA91} noted that most Seyfert galaxies and RQQs obey the
far-infrared/radio correlation $\log(L_\mathrm{FIR}/L_\mathrm{Radio})
\sim -2$ obeyed by nearby star-forming galaxies, while radio-selected
galaxies and RLQs have $\log(L_\mathrm{FIR} / L_\mathrm{radio}) \sim
-5$. If far-infrared/radio luminosity ratios are used to \emph{define}
the RLQ and RQQ populations, then their sample QSOs have a strongly
bimodal distribution in $\log(L_\mathrm{FIR} / L_\mathrm{Radio})$.
\citet{SA91} suggested that the RQQs and Seyfert galaxies are spiral
galaxies whose radio emission is due to disk star formation, while
radio-selected galaxies and RLQs are elliptical galaxies with
SMBH-powered radio emission.  This scenario suggests that radio
sources in RQQs and RLQs are powered \emph{either} by star formation
\emph{or} by AGNs, but rarely by both.  Thus, like optically selected
elliptical galaxies, many QSOs should be ``radio silent,'' with radio
luminosities well below those of normal spiral galaxies.

\citet{KKC11} discovered that the distribution of $\log(S_6)$ has few
sources with $S_6 \lesssim 20 \, \mu\mathrm{Jy}$, a distinct ``bump''
centered on $S_6 \approx 100 \, \mu\mathrm{Jy}$ and a long, flat tail
extending to $S_6 \approx 10^5 \,\mu\mathrm{Jy}$.  This distribution
is not bimodal, but neither is it featureless, so it still suggests
two populations of QSOs---those in the radio-quiet bump and those in
the radio-loud tail.  \citet{CKK13} used NVSS data to statistically
find similar 1.4~GHz flux-density distributions in a much larger
volume-limited sample of 1313 SDSS QSOs with redshifts $0.2 < z <
0.45$ and for a magnitude-limited sample of 2471 bright ($m_\mathrm{r}
< +18.5$) high-redshift ($1.8 < z < 2.5$) QSOs.

In narrow redshift ranges, luminosities are strongly correlated with
flux densities, so the 6 GHz luminosity functions have similar bumps
and tails as the flux density distributions.  Spectral luminosities are more fundamental than observed
flux densities because they are intrinsic source properties
independent of redshift or distance.  \citet{KKC11} and \citet{CKK13}
suggested (but could not prove) that low-redshift ``bump'' radio
sources with 6~GHz luminosities $L_6 \lesssim 10^{\,23}\, \mathrm{W\,
  Hz}^{-1}$ and high-redshift sources with 6~GHz luminosities $L_6
\lesssim 10^{\,24}\, \mathrm{W \, Hz}^{-1}$ are powered more by star
formation in the host galaxy than by the AGN that drives the very
luminous optical and infrared emission, while the more luminous ``tail'' radio sources
are primarily AGN-powered.

In this interpretation, (1) the total radio luminosity of any QSO is
the \emph{sum} of its star-formation and AGN-powered contributions,
and (2) most QSO host galaxies are currently forming stars.  Their
median star-formation radio luminosity ranges from $L_6 \approx
10^{\,22.2}\, \mathrm{W\, Hz}^{-1}$ (about the luminosity of M82) for
low-redshift QSOs just brighter than the $M_\mathrm{B} = -23$ optical
luminosity cutoff to $L_6 \sim 10^{\,23.5}\, \mathrm{W\, Hz}^{-1}$
(about the luminosity of Arp 220) for the most luminous high-redshift
QSOs in the universe ($M_\mathrm{B} \sim -28$).  The relevant measure
of radio loudness is neither the radio flux density $S$ nor the
radio--optical flux density ratio $R$; it is the radio spectral
luminosity $L_\nu$.  The boundary separating RLQs from RQQs is not an
arbitrary flux density or radio--optical flux density ratio; it is the
radio spectral luminosity above which the AGN component usually
dominates and below which the star-forming galaxy component
contributes significantly.  This boundary ranges from $L_6 \sim
10^{\,23} \mathrm{~W~Hz}$ for the least luminous \emph{bona fide} QSOs
($M_\mathrm{B} \lesssim -23$) found at low redshifts to $L_6 \sim
10^{\,24}$ for the most luminous QSOs ($M_\mathrm{B} \sim -28$)
usually found at higher redshifts.  
The radio luminosity function of
the AGN-powered sources alone is so broad and flat that starbursts
contribute more than AGNs to the total radio luminosities of most
QSOs.  If most RQQs are in star-forming galaxies, not just ``red and
dead'' elliptical galaxies, the typical ``radio quiet'' QSO should have
a total radio luminosity slightly higher than that of its host galaxy
alone, and there should be almost no ``radio silent'' QSOs.
Consequently, there is a ``bump'' in the {\it total} radio luminosity
function of QSOs, and that bump peaks just above the typical
luminosity of the star-forming component \citep{KKC11}.

Although some QSOs have intermediate levels of luminosity or $R$,
there is no evidence for a separate class of radio-intermediate
quasars whose radio emission is driven primarily by an SMBH, as
proposed by \citet{FSP96}.  Rather, as described above, we believe
that the radio emission from quasars with $L_{1.4} \sim 10^{24}\,
\mathrm{W\, Hz}^{-1}$ is due to a mixture of AGN and star forming
activity.

Using sub-mJy sources selected from a radio survey of the Chandra Deep
Field \citep{KFM08,MFK08,MBF13}, \citet{PMK11,PBM14,PBK15} have noted
that the radio emission from radio-quiet AGNs is also driven by a
mixture of SMBHs and star formation, and they derived separate
evolving luminosity functions for these two populations. We note,
however, that few of the AGNs identified from these sub-mJy radio
surveys are \emph{bona fide} QSOs with $M_\mathrm{B} < -23$.

Although half a century has passed since the recognition of RQQs,
there is still no consensus whether the RLQs represent a separate
population of QSOs, or whether they are simply the high luminosity
tail of an essentially continuous radio luminosity distribution.
Moreover, if RLQs are indeed a separate population, why are only a
small fraction of QSOs strong radio sources?  A number of
possibilities have been discussed including: \\
\noindent \textbf{Stellar or Black Hole Mass, Accretion Rate, or Black 
Hole Spin}
\citep[e.g.,][]{DMK03,BKH05,SSL07,GMW08,BH12}. 
\citet{WC95} proposed that the main difference
between RLQs and RQQs is associated with black hole spin. Black holes are
spun up by recent galaxy mergers, and the radio jets of RLQs
extract their energy from rapidly spinning black holes.  However, there
is no evidence for black-hole spin powering of jets in X-ray binaries
\citep{FGR10}, and the role of BH spin in powering quasar 
radio jets remains uncertain \citep{HB14}. \\ 
\textbf{Intermittent
  Activity}: The radio emission from quasars varies with
characteristic times scales of months to years.  Typically, the flux
density varies by considerably less than an order of magnitude.
Longer-term periods of more enhanced brightness would be required to
explain the apparent two populations of RLQs and RQQs, and for the
extended radio-loud sources, as suggested by the apparent
double-double quasars \citep [e.g.,][]{JSK09,NRS14}, time scales of at
least $10^5$ to $10^6$ years would be needed.  However, \citet{BLB05}
report that RLQs and RQQs show similar variability on time scales of
months, and they suggest that in both classes of QSOs, the radio
emission is ``intimately associated with the active
nucleus.'' \\
\textbf{Absorption}: Both synchrotron self absorption and
free-free absorption are known to be important in RLQs.  Possibly
absorption could suppress most of the radio emission and give rise to
an apparently radio-quiet population. However, if this were the case
we would expect the division between RQQs and RLQs to be much stronger
at lower frequencies, and there is no evidence for this \citep
[e.g.,][]{MSC12}. \\
\textbf{Host Galaxy Properties}: \cite{DTH93,
  BH12,JKW04,FDK13,FBK14} and others have investigated the host
galaxies of QSOs and found a mixture of early and late type galaxies,
although \citet{DMK03} report that for QSOs with $M_\mathrm{i}<-23.5$
(for $H_0 = 50 \mathrm{~km~s}^{-1} \mathrm{~Mpc}^{-1}$, or
$M_\mathrm{i} < -22.7$ for $H_0 = 70 \mathrm{~km~s}^{-1}
\mathrm{~Mpc}^{-1}$) ``the hosts of both RLQs and RQQs are virtually
all massive ellipticals.''  Thus it appears that the radio loudness of
{\it bona fide} QSOs is not determined by host galaxy
morphology. \\ 
\textbf{Relativistic Doppler Beaming}: Almost all of
the most luminous RLQs exhibit superluminal motion apparently
reflecting relativistic Doppler beaming \citep[e.g.,][]{LCH09}.  As
first suggested by
\citet{SR79}, it is natural to try to interpret RLQs as the subset of
isotropically oriented QSOs whose relativistic beams are aligned close
to our line of sight and so have relativistically boosted radio flux
densities.  However, as discussed in Section~\ref{sec:radioloud}, there
are several problems with this apparently attractive interpretation.

\section{VLA Observations of Low-redshift QSOs in a Volume-limited Sample}
\label{sec:JVLAsample}

In order to avoid the pitfalls encountered by earlier radio studies of
QSOs, as reported by \citet{KKC11}, we selected a complete sample of QSOs 
satisfying  all of the following: \\
(1) SDSS colors well outside the stellar locus in the
$(u-g),\, (g-r), \,(r-i)$ color cube \citep{RFN02}. 
SDSS objects were targeted for spectroscopic follow-up by
 more than one algorithm.  To avoid any bias toward QSO candidates favored
for  spectroscopic confirmation because they were detected by
 the 1.4~GHz FIRST survey \citep{BWH95}, we considered only those QSOs
 chosen on the basis of color. \\
(2) At least one broad emission line with $\mathrm{FWHM~} > 1000
\mathrm{~km~s}^{-1}$ for spectroscopic confirmation of an AGN. \\
(3) $M_\mathrm{i}<-23$ to exclude Seyfert nuclei and other
low-luminosity AGNs.\\
(4) $0.2 < z < 0.3$.  This narrow redshift range minimizes the effects
of evolution and the degeneracy between luminosity and redshift.  The
low maximum redshift $z = 0.3$ ensures that sensitive VLA observations
can detect sources with luminosities only a few times that  of the Milky Way, 
$L_6 \sim 10^{\,21.3}
\mathrm{~W~Hz}^{-1}$. Combined with the requirement that $M_\mathrm{i}
< -23$, $z < 0.3$ also implies $m_\mathrm{i} < 19$ 
after correction for extinction
 according to \citet{sfd}. This is brighter than the SDSS
completeness limit for QSO spectroscopy, so our QSO sample is truly volume
limited. Our sample QSOs are labeled in the SDSS
 quasar catalog with the flag ``low $z$'' \citep{RFN02}, as the 
SDSS targeting algorithm happens
 to be particularly efficient for the identification of low-redshift
 quasars.\footnote{http://classic.sdss.org/dr7/} \\
(5) The $\Omega \approx 3.6$~sr SDSS area
bounded by $90^\circ < \mathrm{~R.A.~}
< 300^\circ$, $b > 30^\circ$ is large enough to yield a statistically
useful number $N = 179$ of color-selected 
QSOs in the \citet{SRH10} quasar catalog of the
seventh data release of the Sloan Digital Sky Survey (SDSS) \citep{AAA09}.  
We later found that one QSO, J1107+080, is badly confused by an
unrelated strong
radio galaxy only $3''$ away in projection but at a much lower redshift
(Section~\ref{sec:sourcenotes}).  As we cannot obtain useful radio
data for J1107+080 and  have no reason to believe that J1107+080 is
intrinsically
different from the other sample QSOs, we excluded it and used only
the remaining 178 QSOs in the following analysis.

We observed using the VLA C configuration with a central frequency of 6
GHz and a bandwidth of 2 GHz in each of the two circular
polarizations.  With natural weighting the synthesized beamwidth was
$3\,\farcs5$ FWHM.

We separated the target QSOs into three groups to optimize the observing time.  About one-third had already been detected as radio
sources by the FIRST \citep{BWH95} and/or the NVSS \citep{CCG98}
surveys with 1.4 GHz peak flux densities $ \geq 1 \mathrm{~mJy}$ per
$5''$ beam or $\geq 2.4 \mathrm{~mJy}$ per $45''$ beam, respectively.
We re-observed these relatively strong sources for only
$90\,\mathrm{s}$ at each of two widely different hour angles.  All
other sources were first observed for $300\,\mathrm{s}$ each.  Those
that were clearly detected with $\mathit{SNR} > 10$ after
$300\,\mathrm{s}$ were observed for $300\,\mathrm{s}$ more at a
different hour angle.  The fainter remaining sources were re-observed
for a total of 30 to 35 minutes, resulting in rms noise levels as low
as $\sigma = 6\,\mu \mathrm{Jy~beam}^{-1}$, which corresponds to a
6~GHz spectral luminosity $L_\mathrm{6} \sim 1.7 \times 10^{\,21} \sim
10^{\,21.2} \,\mathrm{W\,Hz}^{-1} $ at $z = 0.3$.  Observations of the target quasars were interspersed approximately every 20 minutes with a phase and amplitude calibrator.  The flux density scale was determined relative to the source 3C 286 which was observed once each day.

\begin{figure}
%\plotone{Deltas.eps}
\includegraphics[scale= 0.8]{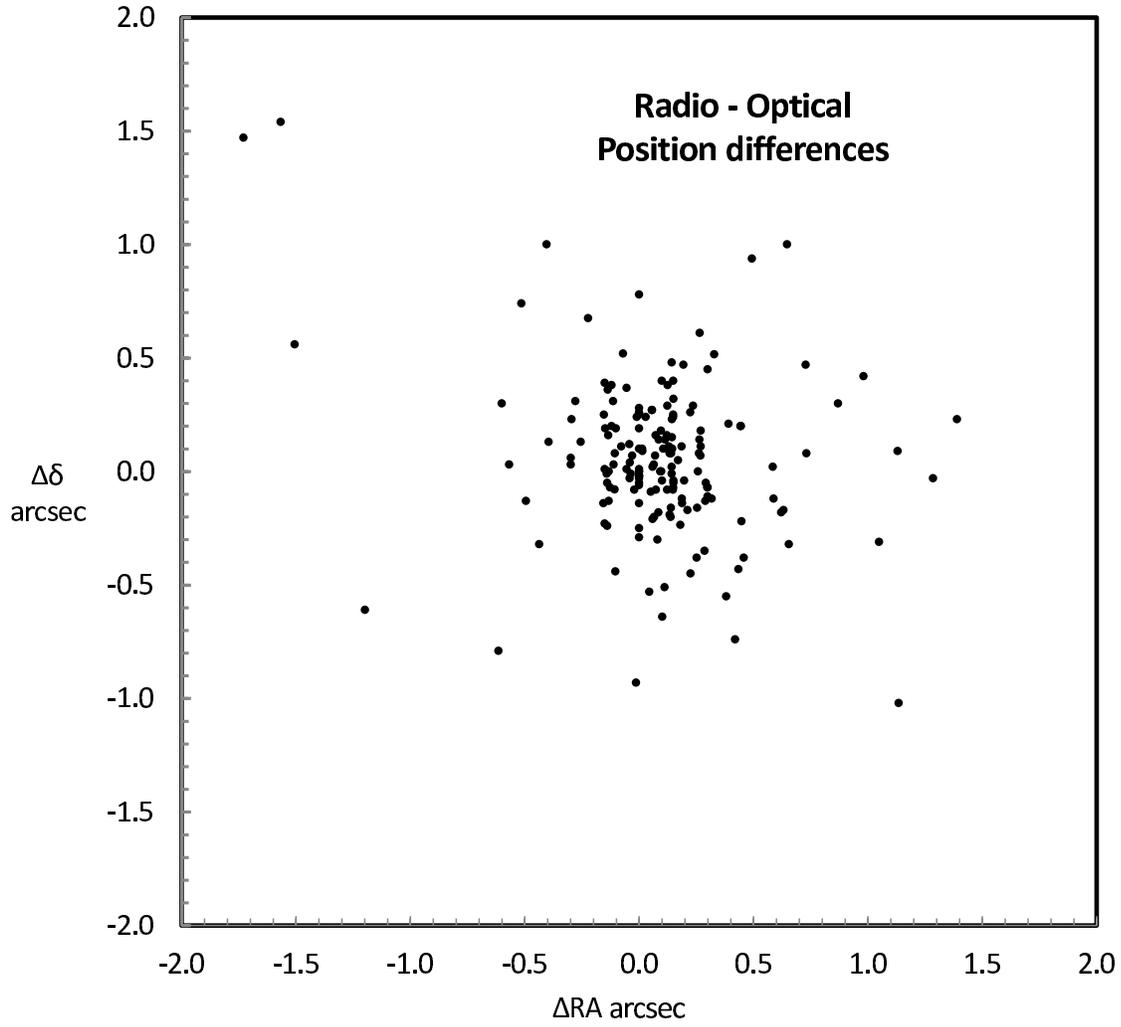}
\caption{Distribution of radio--optical position offsets in arcseconds.  
Not shown are J0919+143 ($\Delta = 8\,\farcs 8$, 29 mJy), 
J1234+644 ($\Delta = 2\,\farcs 5$, $240\, \mu\mathrm{Jy}$),
and J1408+630 ($\Delta = 2\,\farcs 2$, $39\,\mu\mathrm{Jy}$).
\label{fig:radoptoff}
}
\end{figure}

The data were all analyzed within AIPS.  After editing for RFI, the
data were calibrated, the 16 sub-bands combined, then imaged and
cleaned in the AIPS task IMAGR using an 0.2 arcsecond cell size and
natural weighting.  The images typically covered 200 to 400 arcseconds
on a side with resolution of 4 by 5 arcseconds.  Where needed to
suppress the contamination from a strong source elsewhere in the
field, or to enhance the dynamic range of any extended structure, we
did one or more rounds of self-calibration.  We used the AIPS task
JMFIT to measure the peak flux density, angular size, and position of
each compact radio component and the AIPS task IMEAN to measure total
flux densities of complex extended sources.  The systematic position
uncertainties are $\lesssim 0\,\farcs1$ and $\lesssim 0\,\farcs2$ for
the compact radio and optical objects, respectively.  The rms
uncertainty in the radio--optical offset $\Delta$ contributed by noise
alone is $\sigma_\Delta \sim 1.7 \sigma / S_{\mathrm p}$. As shown in
Figure~\ref{fig:radoptoff} in all but two cases, there is an
apparently unresolved radio source near the QSO position.  Although we
describe all flux densities as measured at 6 GHz, because of the very
large bandwidth used, this is strictly true only if the spectral index
is zero.  In the more general case for a source with spectral index
$\alpha$ \citep{C15},
\begin{equation}
\nu_\mathrm{u} = \Biggl[ 
\Biggl( \frac {1} {\alpha+1} \Biggr)
\Biggl( \frac { \nu_\mathrm{max}^{\alpha+1} - \nu_\mathrm{min}^{\alpha+1} } 
{ \nu_\mathrm{max} - \nu_\mathrm{min} } \Biggr) \Biggr]^{1/\alpha} ~, 
\quad (\alpha \neq -1)
\end{equation}
So for a flat bandpass in the range $5 \leq \nu \leq7$ GHz and
$\alpha$ = -0.7, the ``effective'' frequency at which the wideband
flux density equals the actual source flux density at 6.00 GHz is
5.952 GHz, and the reported flux densities at 6.00 GHz are too high by
0.56 percent.  In the extreme case where $\alpha$ = -1, the
``effective'' frequency is 5.94 GHz, and the reported flux densities
too high by 0.94 percent.

\noindent Our results are listed in Table 1 as follows:\\
Column (1): Source name, format SDSS JHHMMSS.SS+DDMMSS.S\\
Column (2): Seconds of right ascension in the VLA position \\
Column (3): Arcsec of declination in the VLA position\\
Column (4): Angular separation $\Delta$ between the radio and optical 
positions (arcsec) \\
Column (5): Redshift from the SDSS\\
Column (6): 6 GHz integrated flux density $S$ ($\mu$Jy) including any extended 
structure and its rms error  calculated as the quadratic sum of the statistical error and a 3 percent uncertainty in the calibration \\
Column (7): 6 GHz peak flux density $S_{\mathrm p}$ ($\mu$Jy~beam$^{-1}$)  
at $3\,\farcs5$ resolution and its rms error calculated as in column 6 \\
Column (8): Ratio of peak to integrated flux density \\
Column (9): Log of the 6 GHz spectral luminosity in the
 source frame (W~Hz$^{-1}$) \\
Column (10): Apparent $i$ magnitude from the SDSS catalog\\
Column (11): Absolute $I$ magnitude\\
Column (12): Ratio $R$ of radio (6 GHz) to optical ($i$ band) flux density
calculated from $S_i = 10^{(9.56-i/2.5)}$~$\mu$Jy\\ 

\clearpage

\begin{deluxetable}{cccccccccccr}
\tabletypesize{\scriptsize}
\rotate
\tablecaption{Results}
\tablewidth{0pt}
\tablehead{
\colhead{(1)} & \colhead{(2)} & \colhead{(3)} & \colhead{(4)} &
\colhead{(5)} & \colhead{(6)} & \colhead{(7)} & \colhead{(8)} &
\colhead{(9)} & \colhead{(10)} & \colhead{(11)} & \colhead{(12)} \\
\colhead{QSO Name} & \colhead{VLA} & \colhead{VLA} & 
   \colhead{$\Delta$} & \colhead{ } & 
\colhead{$S$} & \colhead{$S_\mathrm{p}$} & \colhead{ } & \colhead{$\log(L)$} & 
 \colhead{$i$} & \colhead{$I$} & \colhead{ } \\
\colhead{SDSS} & \colhead{RA (s)} &  \colhead{Dec ($''$)} & 
   \colhead{($''$)} & \colhead{$z$} & 
\colhead{($\mu$Jy)} & \colhead{($\mu$Jy~beam$^{-1}$)} & 
   \colhead{$S_\mathrm{p}/S$} & \colhead{(W~Hz$^{-1}$)} & 
   \colhead{(mag)} & \colhead{(mag)} & \colhead{$R$} 
}
\startdata

J075403.60+481428.0 & 03.617 & 28.05 & 0.18 & 0.276 & $2631   \pm  84.8$  & $2631  \pm 84.8$    & 1.000 & 23.76 & 17.148 & $-23.06$ &   5.24 \\
J080829.17+440754.1 & 29.230 & 55.10 & 1.19 & 0.275 & $61     \pm 11.2$   & $61     \pm 11.2$   & 1.000 & 22.12 & 17.615 & $-22.58$ &   0.19 \\
J081652.24+425829.4 & 52.252 & 29.21 & 0.23 & 0.234 & $232    \pm 13.9$   & $232    \pm 13.9$   & 1.000 & 22.54 & 16.636 & $-23.21$ &   0.29 \\
J082205.24+455349.1 & 05.230 & 49.18 & 0.13 & 0.300 & $222    \pm 11.2$   & $222    \pm 11.2$   & 1.000 & 22.77 & 17.695 & $-22.69$ &   0.73 \\
J083353.88+422401.8 & 53.880 & 01.75 & 0.05 & 0.249 & $209000 \pm 6271$   & $209000 \pm 6270$   & 1.000 & 25.56 & 16.480 & $-23.50$ &   225  \\
\enddata
\end{deluxetable}
\clearpage

\begin{figure}
\includegraphics[angle=0,scale=0.7]{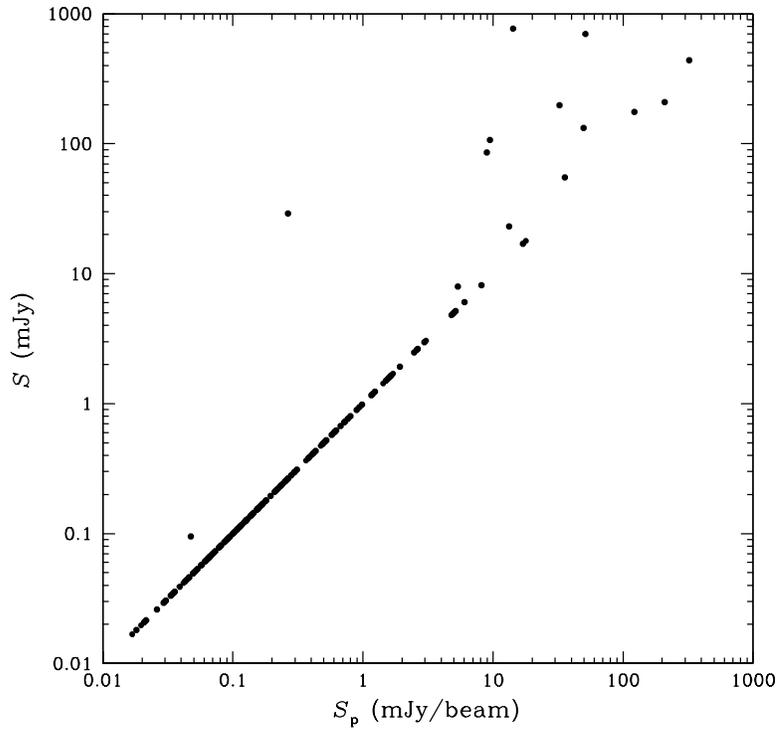}
\caption{
Total flux density $S$ (Column 8 of Table 1) vs. peak flux density 
$S_\mathrm{p}$ (Column 9 of Table 1) for the 178 SDSS QSO in our final 
sample.  The values for 1021+190 and 1034+605 are shown at their 
measured values although they are not considered reliable detections. 
\label{figspsi}
}
\end{figure}

In Figure~\ref{figspsi} we plot the integrated 6 GHz flux densities
$S$ against the peak flux densities $S_\mathrm{p}$.  With one possible
exception, we did not find evidence for any extended structure in any
member of the radio-quiet population.  SDSS J1458+459 is
coincident with an unresolved $43\,\mu$Jy radio source.  A second
source having comparable flux density is located about 5 arcseconds to
the east and close to a fainter un-cataloged optical feature.  It is
not clear if the two radio components are both part of SDSS
J1458+459 or if the eastern radio feature is an independent
source, possibly associated with the un-cataloged optical object,
although there is a $\Delta \sim 1$ arcsecond discrepancy between the
radio and optical positions of the secondary feature.  The other
``discrepant'' point is the radio source J0919+143, which by virtue of
its 29 mJy extended structure is radio loud even though there is only
265 $\mu$Jy in its unresolved core.

\clearpage

\begin{figure}
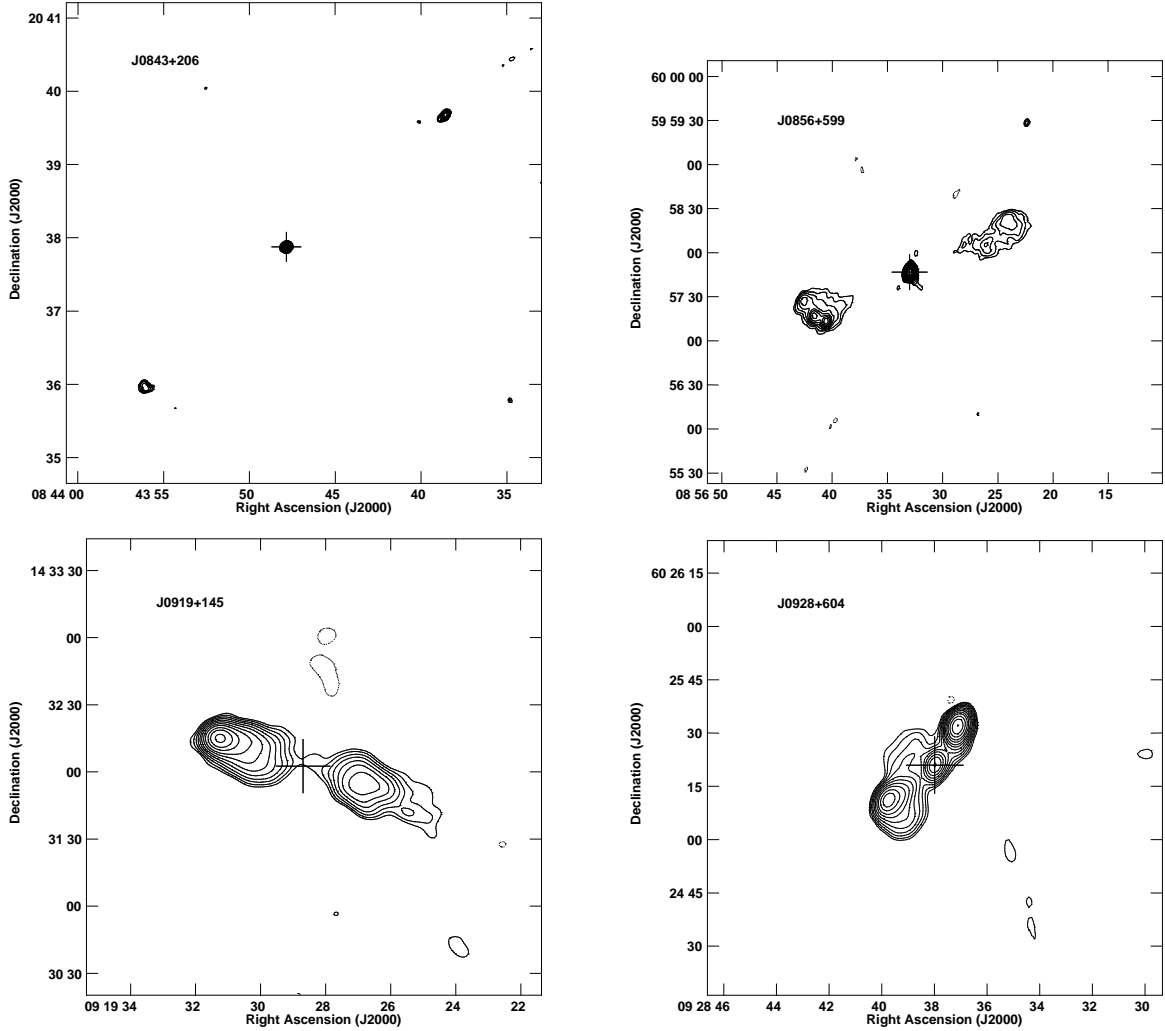

\epsscale{1.0}
\plottwo{0843+206x.ps}{0856+599x.ps} \\
\plottwo{0919+145x.ps}{0928+604x.ps} \\
\caption{Contour maps for the extended radio sources.  The cross 
marks the position of the optical counterpart.  Contour
  intervals are successive factors of $2^{1/2}$ above the lowest contour
  level, which is 
$100\,\mu\mathrm{Jy\,beam}^{-1}$ for J0843+206, 
$200\,\mu\mathrm{Jy\,beam}^{-1}$ for J0856+599
and J0919+145, and
$500\,\mu\mathrm{Jy\,beam}^{-1}$ for J0928+604.
\label{fig:contoura}
}
\end{figure}

\begin{figure}
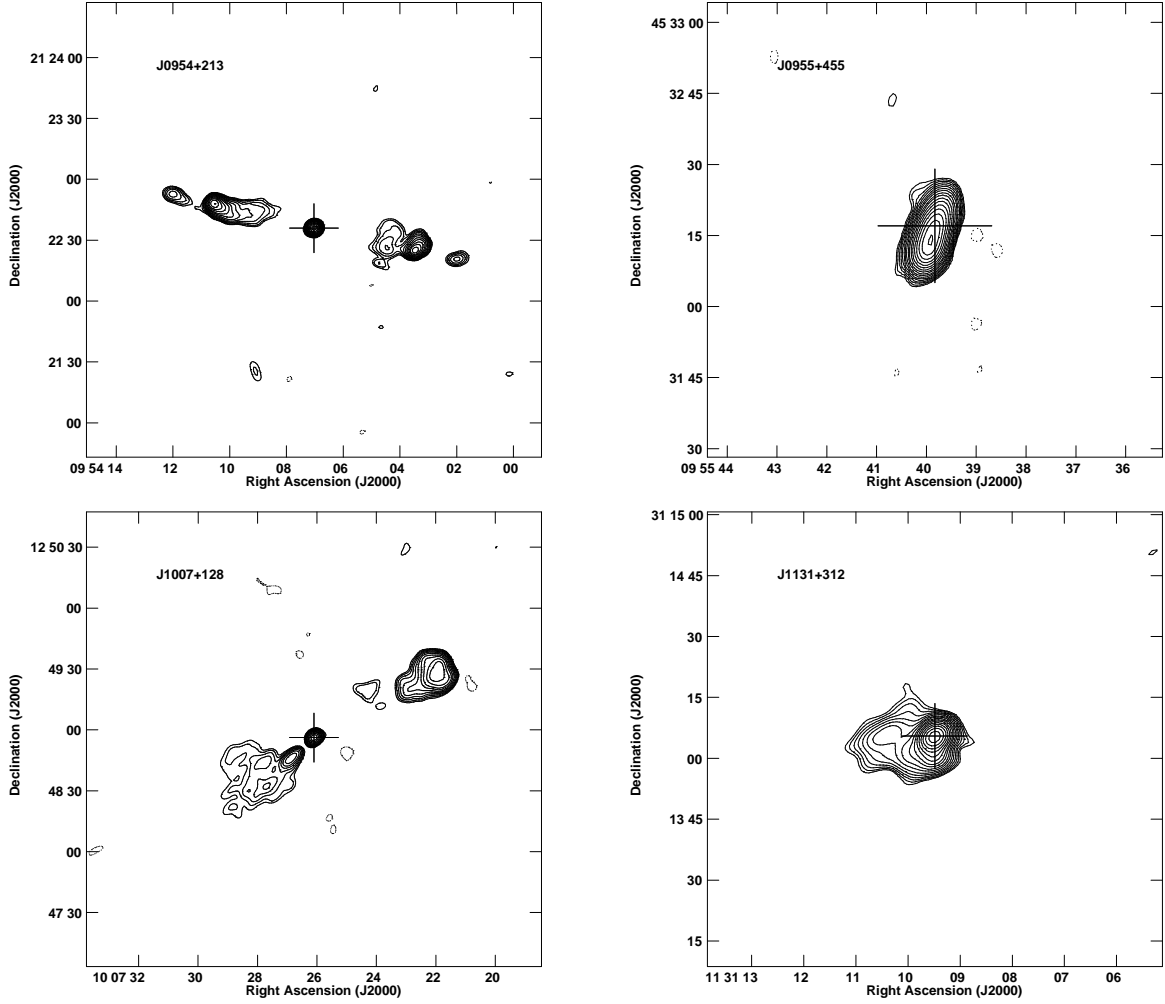

\epsscale{1.0}
\plottwo{0954+213x.ps}{0955+455x.ps} \\
\plottwo{1007+128x.ps}{1131+312x.ps} \\
\caption{Contour maps for the extended radio sources.  The cross 
marks the position of the optical counterpart.Contour
  intervals are successive factors of $2^{1/2}$ above the lowest contour
  level, which is 
$600\,\mu\mathrm{Jy\,beam}^{-1}$ for J0954+213,  
$50\,\mu\mathrm{Jy\,beam}^{-1}$  for J0955+455, 
$500\,\mu\mathrm{Jy\,beam}^{-1}$ for J1007+128, and
$500\,\mu\mathrm{Jy\,beam}^{-1}$ for J1131+312. 
\label{fig:contourb}
}
\end{figure}

\begin{figure}
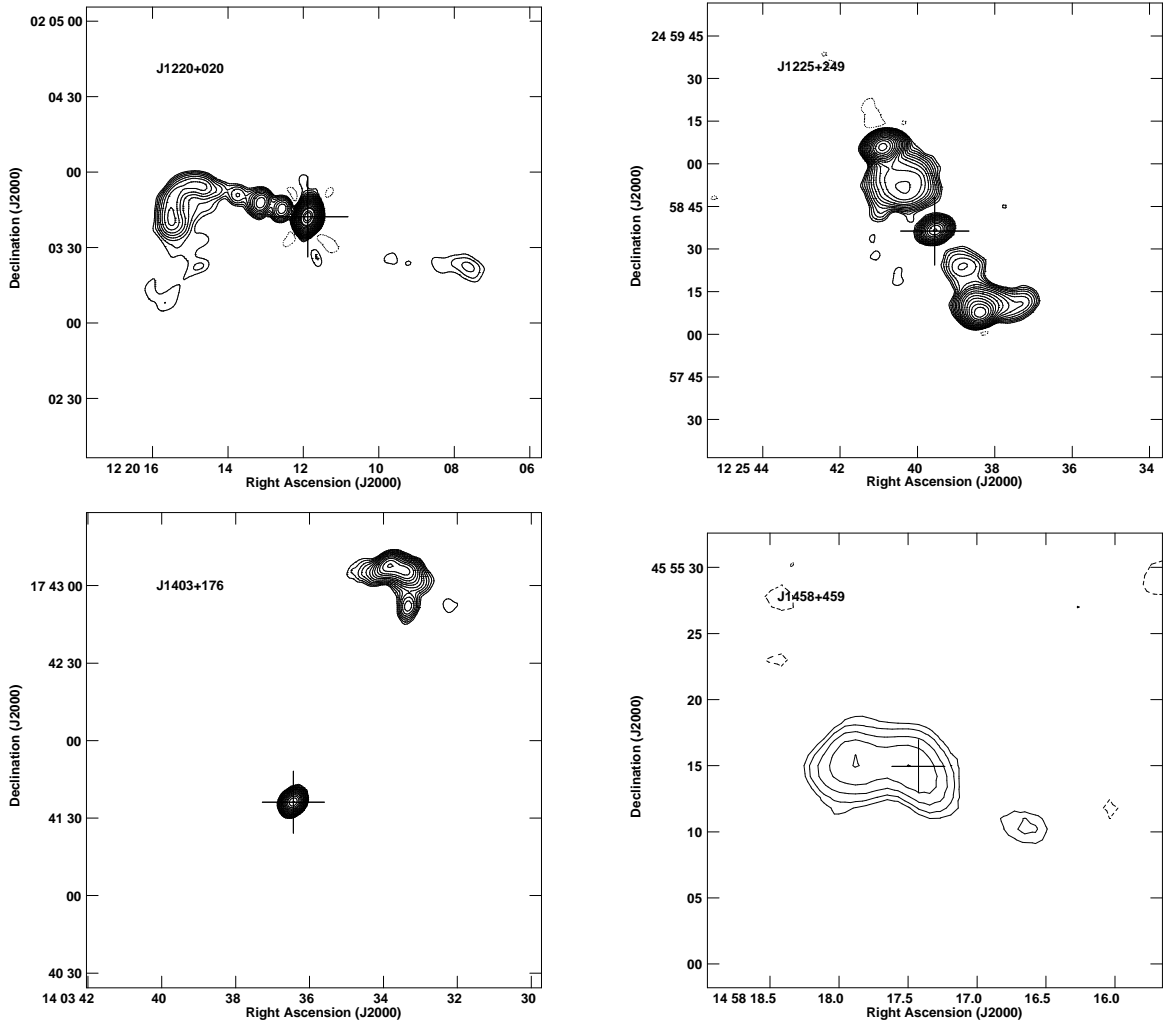

\epsscale{1.0}
\plottwo{1220+020x.ps}{1225+249x.ps} \\
\plottwo{1403+176x.ps}{1458+459x.ps} \\
\caption{
Contour maps for the extended radio sources.  The cross 
marks the position of the optical counterpart.Contour
  intervals are successive factors of $2^{1/2}$ above the lowest contour
  level, which is 
$400\,\mu\mathrm{Jy\,beam}^{-1}$ for J1220+020,
$100\,\mu\mathrm{Jy\,beam}^{-1}$ for J1225+249
$100\,\mu\mathrm{Jy\,beam}^{-1}$ for J1403+176, and 
$10\,\mu\mathrm{Jy\,beam}^{-1}$  for J1458+459.
\label{fig:contourc}
}
\end{figure}

\begin{figure}
\epsscale{1.0}
\plottwo{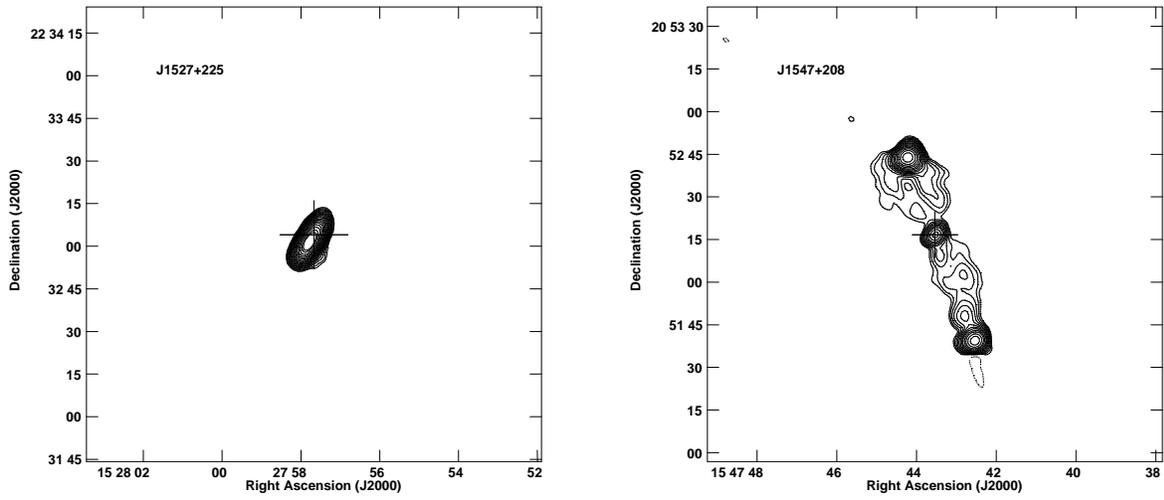}{1547+208x.ps} \\
\caption{
Contour maps for the extended radio sources.  The cross 
marks the position of the optical counterpart.Contour
  intervals are  successive factors of $2^{1/2}$ above the lowest contour
  level, which is 
$200\,\mu\mathrm{Jy\,beam}^{-1}$ for  J152757.67+223304.0
 and
$2000\,\mu\mathrm{Jy\,beam}^{-1}$ for J154743.53+205216.6.
\label{fig:contourd}
}
\end{figure}

Of the 18 SDSS QSOs with unresolved
cores stronger than 5 mJy, seven have no observable extended emission. Another source, J1111+483 has a 4.8 mJy feature located about 2 arcmin northwest of the QSO and which is coincident with a magnitude 21.6 red galaxy; so we consider this feature to be unrelated to the SDSS QSO.  Five RLQs (J0955+455, J1131+312, J1220+020, J1403+176,  and J1527+225)  show one-sided extended structure.  
Seven sources (J0843+206, J0856+599, J0928+604,
J0954+213, J1007+128, J1225+249, and J1547+208) show rather symmetric
double lobes about equidistant from the optical QSO.  One other RLQ, J0919+143 has
prominent extended structure but only weak emission near the
QSO. In general the
extended structure is well resolved, but the outer components of J0843+203
 appear remarkably compact, although there are no SDSS
counterparts for these compact radio components. The only RQQ showing apparent extended structure (J1458+459) has a companion of about equal
flux density located $\approx 5''$ from the QSO that may or may not be
associated with the QSO.
Figures~\ref{fig:contoura}, \ref{fig:contourb}, \ref{fig:contourc},
and \ref{fig:contourd} show contour plots of the 13 RLQ and 1 RQQ with
  extended structure. 

\clearpage

\clearpage

\vfill
\clearpage

\section{Notes on Individual Sourcess}
\label{sec:sourcenotes}

Of 179 SDSS quasars observed, we detected radio emission above the
$3\sigma$ level from all but three. One of them, J1013+020, was
probably detected with flux density $S_6 = 21.5 \,\mu$Jy and
$\mathit{SNR} = 2.5$ at position offset $\Delta = 1\farcs 1$.  The
radio image of J1034+600 has poor sensitivity caused by an $S_6 \sim
100$ mJy source only $2'$ from the target QSO, so it was not detected
with a $3\sigma$ upper limit of 39 $\mu$Jy.  J1021+190 was undetected
with a $3\sigma$ upper limit of 25 $\mu$Jy.  Radio sources near
J1040+600, J1120+427, J1526+279, and J1544+284 had apparent $3 <
\mathit{SNR} < 5$ and offsets $\Delta < 0\,\farcs7$, so we consider
these to be robust detections.  The radio position of J0847+265
differs from the optical position by $\Delta = 1\,\farcs1$, but we
consider this to be an acceptable offset for a source with
$\mathit{SNR} = 3.4$.  For J0919+145, the formal position discrepancy
is $\Delta = 8\,\farcs8$, but J0919+145 is a strong double-lobe source
with a component separation of $\sim 90''$
(Figure~\ref{fig:contoura}).  Its $265 \, \mu \mathrm{Jy}$ compact
core contains only about 1\% of the total flux density $S_6 = 29
\mathrm{~mJy}$ and is confused by the strong western lobe, so we do
not consider the formal measured radio--optical position difference to
be significant.

The radio position of the relatively strong 242 $\mu$Jy source
J1234+644 differs from the SDSS position by $\Delta = 1\,\farcs5$.
Inspection of the SDSS image reveals a nebulosity surrounding the QSO
which may be the QSO host galaxy.  Superimposed at the approximate
position of the radio source is a condensation which may be part of
the host galaxy or possibly a faint foreground galaxy.

The strong ($S_6 = 4.9 \mathrm{~mJy}$) radio source J1107+080 is
offset from the QSO position by $\Delta = 3\,\farcs05$ and coincides
($\Delta < 0\,\farcs2$) with an unrelated $i = 15.1$ galaxy at redshift $z =
0.0734$.  The angular separation is comparable with our HPBW, so we
cannot make a meaningful measurement of the flux density of the QSO.
Therefore we removed this source from our final sample, which reduces
it from 179 to 178 QSOs.

We consider all but J1034+605 and J1021+190 to be robust radio
detections of radio emission from SDSS quasars. Even if one of the
low-$\mathit{SNR}$ detections is not real, the corresponding SDSS
QSO is still radio quiet, with maximum flux density well below the
$S_6 \approx 100 \,\mu \mathrm{Jy}$ peak of the radio-quiet
distribution, and our conclusions are unchanged.

\section{Two QSO Populations}
\label{sec:twopopulations}

\begin{figure}
\epsscale{1.1}
\plotone{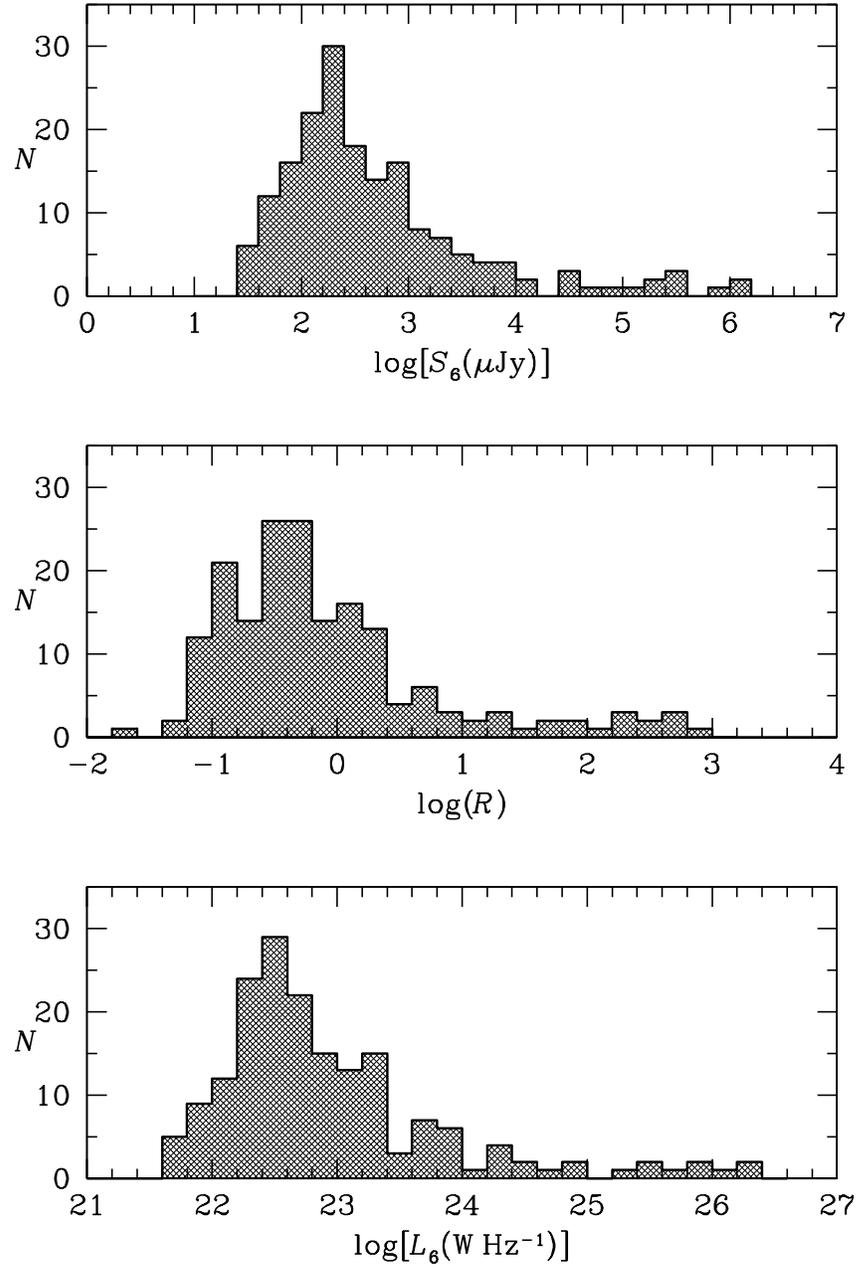}
\caption{ Distributions of the radio flux densities, radio/optical
  flux-density ratios, and radio luminosities from Table 1 columns 6,
  12, and 9, respectively.  Quasars SDSS J1021+190 and SDSS
  J1034+608 are plotted at their observed values although
  they are only marginal detections.
\label{fig:SLRhisto}
}
\end{figure}

\begin{figure}
\epsscale{1.1}
\plotone{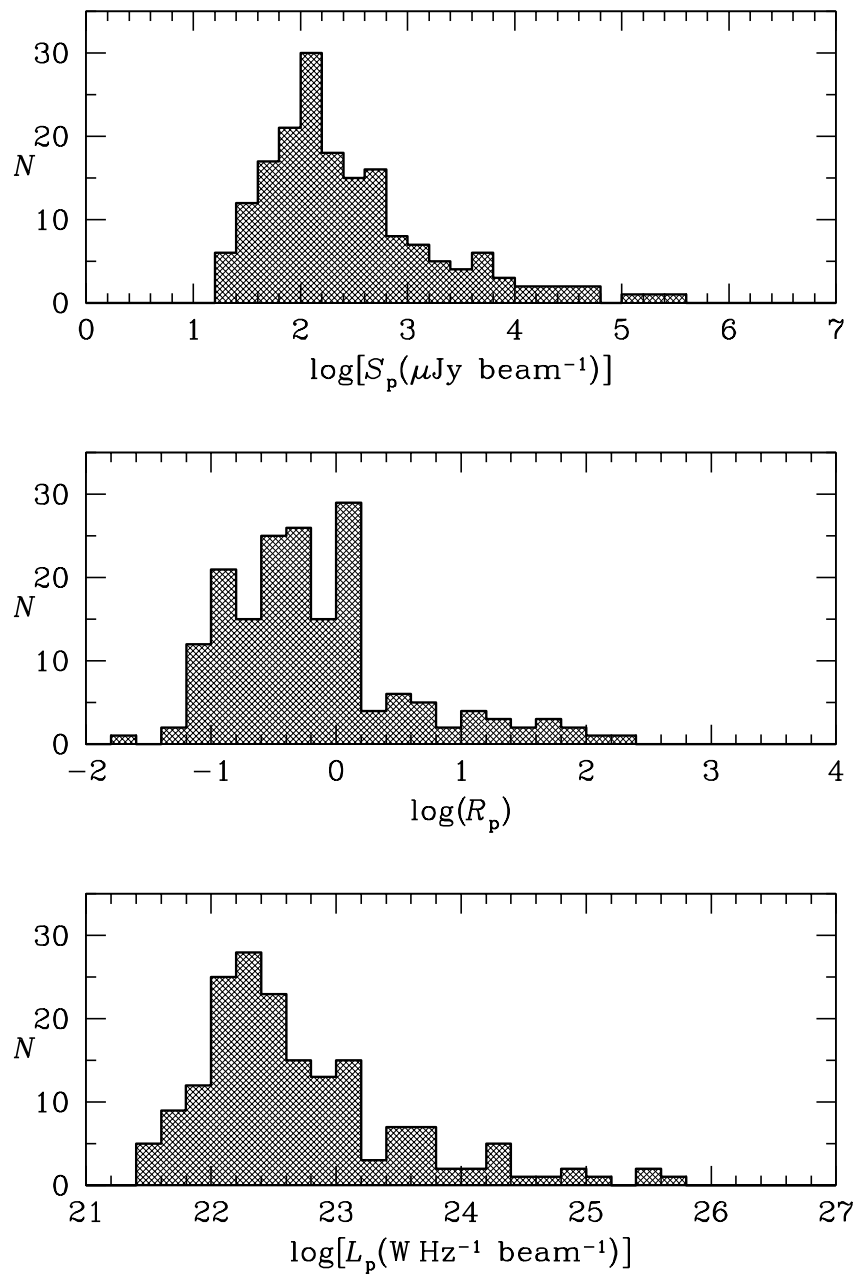}
\caption{ Distributions of the radio peak flux densities (column 7 of
  Table 1), the corresponding peak radio/optical flux-density ratios,
  and the luminosities calculated from the peak flux densities.
  Quasars SDSS J1021+190 and  SDSS J1034+608 are
  plotted at their observed values although they are only marginal
  detections.
\label{fig:SLRpeakhisto}
}
\end{figure}

In Figure~\ref{fig:SLRhisto} we show the logarithmic distributions of
measured flux densities $S_6$, the ratios $R$ of radio (6 GHz) to
optical ($i$ mag) flux densities, and the 6 GHz spectral luminosities
$L_6$ for all 178 sources remaining in our SDSS
sample. Figure~\ref{fig:SLRpeakhisto} shows the same distributions
based on peak flux densities in the $3\,\farcs5$ beams coincident with
the QSOs.

In each of these distributions, we clearly see the presence of two
populations as reported earlier by \citet{KKC11} and
\citet{CKK13}. One has 6 GHz spectral luminosities $L_6 \lesssim
10^{\,23}\, \mathrm {W\, Hz}^{-1}$. We suggest that the radio emission from these RQQs is the result of
active star formation in the host galaxies, and the other more
luminous but much smaller population is probably directly related to
the central engine associated with the SMBH which powers the optical
luminosity.  We note that the distributions in
Figures~\ref{fig:SLRhisto} and \ref{fig:SLRpeakhisto} are very
similar, so the radio-loud population cannot result from the extended
emission alone.

\begin{figure}
%\epsscale{1.1}
%\plotone{EVLArhom2.eps}
\includegraphics[scale=1., clip=true, trim=0.5in 1.5in 1.in 2.in]{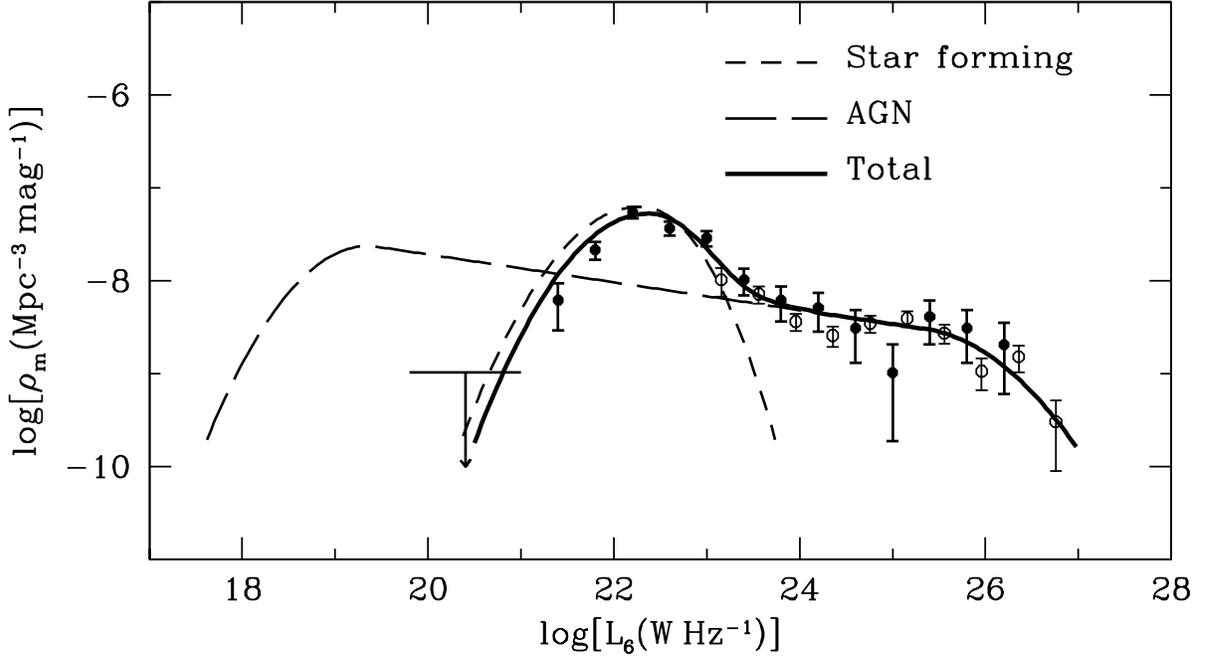}
\caption{The total 6~GHz radio luminosity function of low-redshift
  QSOs (solid curve) fitted to our 6 GHz VLA data (filled circles and
  upper limit representing the two undetected sources) and NVSS
  \citep{CCG98} data converted from 1.4 GHz to 6 GHz \citep{CKK13}
  using $\alpha = -0.7$ (open circles) can be decomposed into
  component luminosity functions contributed by their star-forming
  host galaxies (short dashed curve) and their AGNs (long dashed
  curve).  At low luminosities the solid curve lies below the dashed
  curves because the total radio luminosity of each QSO is greater
  than that of either component.
\label{fig:RLF6}
}
\end{figure}

We also recalculated the 6 GHz QSO luminosity function in the redshift
range $0.2 < z < 0.3$ using our new flux densities and upper limits.
It is shown in Figure~\ref{fig:RLF6} and is similar to Figure~2 in
\citet{KKC11}, the most obvious difference being the tighter upper
limit to $\log(\rho_\mathrm{m})$ below $L_6 \sim 10^{\,21}
\mathrm{~W~Hz}^{-1}$ because four of the six formerly undetected QSOs
were detected in the reprocessed data.

\clearpage

\subsection{The Radio Quiet QSO Population}\label{sec:radioquiet}

The 6 GHz radio luminosity $\log[L_6 (\mathrm{W~Hz}^{-1})] \approx
22.4$ of the RQQ peak (solid curve in Figure~\ref{fig:RLF6}) and the
median spectral index $\langle \alpha \rangle \sim -0.7$ \citep{CKK13}
of low-redshift RQQs are characteristic of radio emission by
relativistic electrons accelerated in supernova remnants of young
stars, and they differ from the flat-spectrum optically thick sources
almost exclusively associated with AGN. Our hypothesis that the radio
emission from RQQs is primarily due to star formation in the host
galaxy and is unrelated to the SMBH \citep{CKK13} is supported by
\citet{JKW04}, who reported evidence for ongoing star formation even
in the elliptical host galaxies of QSOs.  Radio emission from galaxies
with active star formation is tightly correlated with FIR emission and
is characterized by the ratio $q \equiv \log(S_\mathrm{FIR}/S_{1.4})
\sim 2.3$ or $\log(S_\mathrm{FIR}/S_6) \sim 2.7$.  The 6 GHz radio
luminosity function that we attribute to star formation (short dashed
curve in Figure~\ref{fig:RLF6}) peaks at $\log[L_6
  (\mathrm{W~Hz}^{-1})]\approx 22.3$ and corresponds to a
star-formation rate $\dot{M} \sim 20 \, M_\odot
\mathrm{\,yr}^{-1}$\citep{CKK13}.

We note that the absolute radio luminosities of our RQQs are far below
the sensitivity limits of nearly all published QSO studies, which have
higher flux-density limits and/or include QSOs with higher redshifts.
They are also far below the traditional
$\log[L_{8.4}(\mathrm{W~Hz}^{-1})] \approx 25$ boundary often used to
distinguish RLQs from RQQs \citep{HIF96}.

\begin{figure}
\includegraphics[scale=1., clip=true, trim=1.in 4.in 0 0]{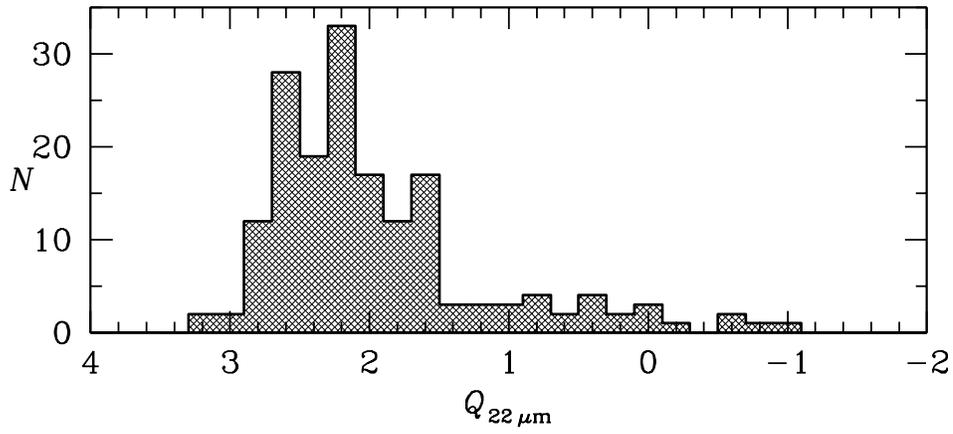}
\caption{ Distribution of the ratio of MIR ($\lambda =
  22\,\mu\mathrm{m}$) to peak radio (6 GHz) flux density.  The sources
  J1021+190 and J1034+608 are shown at their measured values
  although they are not considered significant detections.
\label{fig:mirradio}
}
\end{figure}

As we do not have FIR data for our SDSS sample, following
\citet{SSM10}, \citet{BBP13}, and others, we have looked at the
distribution of $Q_{22} \equiv \log (S_{{22}\,\mu\mathrm{m}}) / S_6)$
(Figure~\ref{fig:mirradio}) based on Band 4 WISE observations and our
6 GHz VLA data.  While Figure~\ref{fig:mirradio} does reflect the
separation between the radio-loud and radio-quiet populations, we do
not see the strong correlation between the $22\,\mu\mathrm{m}$ and 6
GHz flux densities that is characteristic of radio - FIR comparisons.
This is not surprising as the measured $22\,\mu\mathrm{m}$ flux
densities are probably dominated by warm dust heated by the AGN
\citep{PMC10} and so are not a clean measure of star-forming activity
in QSOs.

In principle, radio emission resulting from star formation may be distinguished from black hole AGN emission by examining the radio spectral index.  For the radio loud sample, we have compared our 6 GHz images with the 1.4 GHz FIRST and NVSS images. As is characteristic of radio loud quasars and radio galaxies, our radio loud sample is a mixture of flat spectrum cores and steep spectrum extended structure consistent with an AGN origin, but there is no corresponding data for the RQQs at other wavelengths.   Ideally one might determine radio spectra of the RQQs just using our 16 separately calibrated sub-bands which cover the frequency range from 5 to 7 GHz. However, as shown by \cite{C15} (eqn. 52), the uncertainty in spectral index calculated in this way is $\sim$ 10/SNR.  As seen from figure 7, the typical RQQ has a flux density $\sim$ 100 microJy, where the  SNR is about 15. Thus for the typical RQQ the rms uncertainty in the spectral index would be $\sim$1, or not very useful to discriminate between a star formation or AGN origin.  

Recently \citet{ZG14}  and \citet{ZLP16} have proposed that the radio emission from many RQQs
is synchrotron radiation from relativistic electrons that are
accelerated by moving shocks in organized quasar outflows, rather than
by star formation.  Their proposal was motivated by their observed
correlation between radio luminosity and outflow velocities measured
from [OIII] $\lambda5000$ \AA~emission-line blueshifts in a sample of
568 obscured luminous Type II quasars in the SDSS.  However, 386 of
the 568 quasars were detected with $S_{1.4} \gtrsim 1 \mathrm{~mJy}$
by FIRST \citep{BWH95}.  By our criterion, this is a sample dominated
by radio-selected quasars, not optically selected QSOs.  Their median
spectral luminosity is $\nu L_\nu = 10^{40} \mathrm{~erg~s}^{-1}$ at
$\nu = 1.4 \mathrm{~GHz}$, or $L_6 \sim 3 \times 10^{23}
\mathrm{~W~Hz}^{-1}$ for sources with $\alpha = -0.7$. This is well
above the radio luminosites of sources that we suggest are powered by
star formation in the host galaxies of low-redshift QSOs
(Figure~\ref{fig:RLF6}), and it would indicate an ``astonishing''
$\dot{M} \sim 400 \, M_\odot \mathrm{~yr}^{-1}$ star-formation rate
\citep{ZG14}.  We agree that the radio emission from these quasars is
not starburst-powered, but  their quasar population does not overlap our RQQ
population.

\cite{HR16} and \cite{MPN16} have detected milliarcsecond structure in 5  RQQs, which they argue is evidence for an AGN driven basis for radio emission in RQQs.  Their radio sources have a 1.4 GHz luminosity greater than $\sim$ $L_{1.4} = 10^{24} \mathrm{~W~Hz}^{-1}$, the value that is characteristic of RQQs at redshifts $\sim$ 1.5 \citep{CKK13}.  However, we do not exclude the possibility that there is a mixture of AGN and star forming activity that contributes to the high end of the RQQ luminosity function. 

\citet{WJH15} found a ``bump'' peaking near $S_{1.4} = 100~\mu
\mathrm{Jy}$ in the normalized QSO count $S^2 n(S)$ contributed by 74
faint ($18.4 \leq K_\mathrm{S} \leq 22.4$ and $M_i \leq -22$)
higher-redshift ($0.7 \lesssim z \lesssim 2.8$) VIDEO QSOs.  They
concluded that this radio emission is predominantly powered by
accretion onto SMBHs.  Again, the typical radio luminosity in this
bump is so high that we do not expect that it can be explained by star
formation in low-luminosity QSOs. Our criterion that RQQs have
radio spectral luminosities low enough to be explained by star
formation ($L_6 \lesssim 10^{23} \mathrm{~W~Hz}^{-1}$) excludes most
QSOs that have been called RQQs in the literature.

\subsection{The Radio Loud QSO Population}\label{sec:radioloud}

Thirteen RLQs 
 show extended radio structures, all but perhaps two of which are spread
over some tens of arcsec or $\sim 100$ ~kpc, or much larger than the host galaxy of stars.  Thus it is clear that at least this large scale structure is due to the AGN and can not, as suggested by Terlivich and collaborators, be due to star formation.

Following the widespread recognition of relativistic beaming as the cause of apparent superluminal motion and rapid flux density variability in RLQs, \citet{SR79} made the elegant suggestion that RLQs are just the subset of quasars whose relativistic beams are oriented nearly along the line of sight.  Previous radio observations of optically selected QSO samples appeared inconsistent with this simple beaming interpretation.  However,
the limited sensitivity of these early observations combined with
poorly defined QSO selection criteria, use of inappropriate
statistics, and differing criteria for separating RLQs from RQQs
led to quantitative disagreements with the \citep{SR79} model. Specifically, it appeared that too many optically selected QSOs appeared radio loud and that  the observed detection rate as
a function of decreasing flux density was not consistent with simple beaming models.

In view of
recent improved observational studies and our increased understanding
of relativistic beaming
\citep[e.g.,][]{CLH07,LCH09,HLK15}, 
we decided to revisit
relativistic beaming models to see if the observed
radio-loud\,/\,radio-quiet dichotomy for the parsec-scale emission can
be explained simply by orientation effects.  Following \citet{SR79},
in Appendix A we derive the expected  radio source count for a
population of randomly oriented relativistic jets assuming that the
optical emission is unbeamed.  As shown by Equation~\ref{eqn:nsslopes}, the
differential number $n(S)\,dS$ of detected radio sources between $S$
and $S+dS$ follows a simple a power law $n(S) \propto S^{\,\zeta}$
whose exponent $\zeta$ depends only slightly on the spectral index and
geometry of the relativistic beam.  For realistic values of spectral
index and geometry $-2 < \zeta < -5/4$, so the number of detected
radio sources is expected to increase moderately with decreasing flux
density.

Considering only the compact
components (Column 9 of Table 1) of the radio loud sources alone, we
found a best fitting \citep{CJM70} power law index of $-1.7 \pm 0.2$
for the differential source count based on the 18 SDSS sources stronger than 5 mJy.  This
is marginally steeper than expected from beaming models with flat
($\alpha \sim 0$) spectra which are characteristic of compact
blazars.  However, our sample is small, and due to the small redshift and corresponding small volume of our SDSS sample, the radio luminosities are low compared with the more powerful, but more rare RLQs, so we would not expect beaming to be very strong in these sources.

Using 1.4 GHz NVSS data for 163 SDSS QSOs stronger than 2.4 mJy and in
the redshift range 0.2 $< z < 0.45$, \citet{CKK13} reported $\zeta
=-1.20 \pm 0.02$, which is consistent with  beaming
models.  For a sample of 191 more luminous optically selected quasars
in the flux-density range $2.4 < S\,(\mathrm{mJy}) < 1000$, and redshift
range $1.8 <z <2.5$, \citet{CKK13} determined $\zeta = -1.01 \pm
0.02$, which is too large; that is, the observed source count is too
``flat'', increasing too slowly as the flux density decreases.  The $3
\sigma$ lower limit $\zeta > -1.07$ is consistent with
Equation~\ref{eqn:nsslopes} only for unreasonably steep spectral indices
$\alpha \lesssim -10$.  In each case, the quoted uncertainties refer to the
statistical uncertainty only.  Complications such as oblique shocks, a
range of intrinsic luminosities or component velocities, contamination
by unbeamed lobe emission, etc. only make it harder for beaming to
match the data. 

Another early objection to the beaming model arose because of the
unexpected large fraction of optically selected QSOs that appeared to
be radio loud.  Because the radiation is beamed into a solid angle
$\sim 1/\gamma^2$, with typical observed values $\gamma \sim 10$, it
was argued that less than 1 percent of optically (randomly oriented)
QSO should appear radio loud, whereas most investigations found 10 to
15 percent to be radio loud.  However, in estimating the fraction of
optically selected QSOs expected to be radio loud QSOs, the assumption
of typical Lorentz factors $\sim 10$ may not be appropriate.  Although
samples of radio selected blazars may have typical Lorentz factors
$\sim 10$, as shown by \citet{LM97} and \citet{LCH09}, the
corresponding much larger parent population is dominated by sources
with smaller Lorentz factors and corresponding larger beaming angles.
Furthermore, the number flux density distribution predicted from
beaming models assumes a simple model with a straight and an
infinitely thin jet with a pattern speed equal to the particle
speed. However, none of these simplifying assumptions are precisely
met in practice \citep{V95,LCH09,HLK15}, particularly the assumption
of straight jets without bends. Also, we note that if the optical
emission is also beamed, then even optically selected QSOs will be
favorably oriented along the line of sight.  Finally, in the so called
unified models, an obscuring torus may block optical and IR radiation
for those QSOs which are oriented close to the plane of the sky, so
that even without consideration of optical beaming, the orientation of
even optically selected QSOs may be biased to favor orientations close
to the line of sight \citep{B89}.

The extended radio emission seen in more than half of our RLQs also represents
a clear challenge to the \citet{SR79} relativistic beaming
interpretation of the radio-loud\,/\,radio-quiet dichotomy. The five RLQs with
asymmetric large-scale structure might be accommodated within modest
relativistic beaming models, and indeed such an interpretation is
supported by the facts that parsec and kiloparsec scale structures of RLQs 
are generally oriented in the same direction \citep{KPT81,B89} and
that RLQs with kiloparsec-scale structures tend to have somewhat
slower measured superluminal velocities in the corresponding
parsec-scale structures \citep{Z97}.  However, the nearly half of the RLQs with two-sided extended structure implies little or no differential Doppler
boosting suggesting that the orientation of Doppler
boosting is not the major effect separating the RLQs from the RQQs.

As shown in Figure~\ref{fig:SLRpeakhisto} the distributions of peak
(parsec scale) flux densities $S_\mathrm{p}$, radio-optical ratios
$R$, and radio luminosities $L_6$ are all similar to the corresponding
distributions derived from total flux densities.  Thus it is clearly not the
extended emission alone that distinguishes RLQs from RQQs.

\section{Summary}\label{sec:summary}

We have reprocessed the VLA observations of a sample of SDSS QSOs discussed in \citet{KKC11}, and generated a catalog of radio sources associated with each QSO.  We confirm  the idea that QSO radio emission comes in two
flavors, as indeed some have argued for decades, and strengthen the suggestion made in \citet{KKC11} that the emission in most
luminous objects is associated with SMBHs, while the emission in the less
luminous objects that dominate the sample is associated with the
formation of stars in the host galaxies.
We detected radio emission at 6 GHz from all but two of the 178
color-selected SDSS QSOs contained in our volume-limited sample of
QSOs more luminous than $M_i = -23$ and with redshifts $0.2 < z <
0.3$ .

About 20\% of these QSOs have 6 GHz source-frame luminosities $L_6
\gtrsim 10^{23}\,\mathrm{W\,Hz}^{-1}$, high enough that SMBHs are
probably needed to power most of their radio emission; we call these
radio-loud QSOs (RLQs).  The RLQ radio luminosity function is flat and
featureless. Extrapolating it to lower luminosities indicates that the
radio power from many SMBHs should be much less than $L_6 \sim 10^{21}
\mathrm{~W~Hz}^{-1}$.  Radio-quiet QSOs (RQQs) cause a ``bump'' in the
QSO luminosity function spanning $21 \lesssim
\log[L_6(\mathrm{W~Hz}^{-1})] \lesssim 23$, and ``radio-silent'' QSOs
fainter than $L_6 \sim 10^{21} \mathrm{~W~Hz}^{-1}$ are actually quite
rare, accounting for not more than 1\% of all QSOs.  We suggest that
the RQQ ``bump'' is evidence for radio emission powered by star
formation in QSO host galaxies, and the lack of radio-silent QSOs is
evidence that most RQQ host galaxies are forming stars at least as
fast as the the Milky Way.  Thus RLQs and RQQs may represent two
separate populations whose radio emission is dominated by distinctly
different energy sources. These populations can be distinguished only
by very sensitive radio observations of low-redshift QSOs.  Nearly all
published attempts to distinguish such RLQs from RQQs have not reached
the necessary luminosity level, $L_6 \lesssim 10^{\,21}
\mathrm{~W~Hz}^{-1}$.

Many of the RLQs show extended structures, most of which are rather
symmetrically located about the parent QSO.  While relativistic
beaming may play a role in the determining the parsec-scale radio flux
densities of some QSOs, the presence of RLQs with symmetric extended
structures would seem to require that another mechanism as well is
important in determining whether or not a QSO becomes radio
loud. Moreover, the flux-density distribution of the RLQ population does
not seem to be consistent with simple relativistic beaming models.

After 50 years of studies of QSO radio dichotomy, technology has
finally advanced enough to provide clues for the behavior of
radio-quiet objects.

\acknowledgments

The VLA is a facility of the National Radio Astronomy Observatory
which is operated by Associated Universities, Inc., under a
Cooperative Agreement with the National Science Foundation. 
Part of this work was done while one of us (AK) was at the CSIRO
Astronomy \& Space Science Australia Telescope National Facility.  We thank Paolo Padovani and the referee for comments and suggestions which have improved the presentation.

{\it Facilities:} \facility{VLA}.

\appendix
\section{Appendix}

When a single source approaches an observer with velocity $\vec{v}$
at an angle $\theta$ from the line of sight,
the observed frequency $\nu$ equals the emitted frequency $\nu'$ multiplied 
by the Doppler factor
\begin{equation}
\label{eqn:Dopplerfactor}
\delta \equiv \frac{\nu}{\nu'} = 
[\gamma (1 - \beta \cos\theta)]^{-1} = 
[\gamma (1 - \beta l)]^{-1}~,
\end{equation}
where $\beta \equiv v / c$, $\gamma \equiv (1 - \beta^2)^{-1/2}$ is the
Lorentz factor, and $l \equiv \cos\theta$ is the direction cosine of 
$\theta$. 

For a population of sources with
randomly oriented velocities, the probability that $\vec{v}$ will point
toward any solid angle $\Omega$ is $\Pi (\Omega) = (4 \pi)^{-1}$, and
the probability distribution of Doppler factors $\delta$ 
can be calculated from
$ \Pi (\delta) \,d \delta =  \vert \Pi (\Omega) \,d \Omega \vert$. 
 \begin{equation}
\label{eqn:pofdelta}
\Pi (\delta) =  \frac {1}{4 \pi} \bigg|\frac{d \Omega}{d \delta} \bigg|
= \frac{1}{4 \pi} \bigg| \frac{d \delta}{d l} 
\frac{d l}{d\Omega} \bigg|^{-1}
\end{equation}
The ring between $\theta$ and $\theta + d \theta$ covers solid angle
\begin{equation}
\label{eqn:dOmega}
d \Omega = 2 \pi \sin \theta \,d \theta = - 2 \pi \,d (\cos \theta)
= -2 \pi dl
\end{equation}
and
\begin{equation}
\label{eqn:ddeltadl}
\frac{d \delta} {d l} = \frac{\gamma \beta} 
{[\gamma ( 1 - \beta l)]^2} = \gamma \beta \delta^2
\end{equation}
so
\begin{equation}
\label{eqn:pofdeltafinal}
\Pi (\delta) = (2 \gamma \beta \delta^2)^{-1}
\end{equation}
in the range
\begin{equation}
\label{eqn:deltarange}
[\gamma(1 + \beta)]^{-1} < \delta < [\gamma(1 - \beta)]^{-1}
= \gamma (1 + \beta)
\end{equation}
corresponding to the range of angles $0 < \theta < \pi$.  In the
limit $\beta \rightarrow 1$,
\begin{equation}
\label{eqn:maxdeltarange}
(2 \gamma)^{-1} < \delta < 2 \gamma~.
\end{equation}

The
quantity $I_\nu / \nu^3$, where $I_\nu$ is spectral brightness,
 is Lorentz invariant
\citep{ryb79}.
The observed flux density $S(\nu)$ 
of a moving source emitting flux density $S'(\nu' )$
isotropically in its frame is
\begin{equation}
\label{eqn:dopplerboost}
\frac{S(\nu)} {S'(\nu')} = \biggl( \frac {\nu} {\nu'} \biggr)^3 = \delta^3~.
\end{equation}
\citep{ryb79}.  In terms of the spectral index $\alpha \equiv + d \ln S / d \ln \nu$
(positive sign convention), the flux density of a source with a power-law
spectrum is
\begin{equation}
\label{eqn:powerlaw}
S'(\nu') = S'(\nu) \biggl( \frac{\nu'}{\nu} \biggr)^\alpha = \delta^{-\alpha}
\end{equation}
so
\begin{equation}
\label{eqn:boostflux}
S(\nu) = S'({\nu})\, \delta^{3 - \alpha}
\end{equation}
Equation~\ref{eqn:boostflux} applies to a single emitting region, or
source component.  The observed lifetime $\tau$ of a source with
proper lifetime $\tau'$ is $\tau = \delta^{-1} \tau'$.  The
time-averaged flux density of a radio source consisting of multiple
identical components with the same velocity
created at a nearly unifom rate $\gg \tau^{-1}$ (e.g., a radio jet)
will be proportional
to $\tau$, or $S(\nu) \propto \delta^{2 - \alpha}$.  To allow for time averaging
over a population of
source components having finite lifetimes and different Doppler factors,
let
\begin{equation}
\label{eqn:observedflux}
S(\nu) = S'(\nu) \, \delta^{\,p - \alpha}~,
\end{equation}
where $S'(\nu)$ is the total flux density in the source frame
and $p = 2$.  Again, for any single component, $p = 3$.

Doppler boosting broadens the flux-density distribution of a
population of intrinsically identical sources characterized
by the same $\gamma$ and $S'(\nu) =
1$.  The probability distribution of observed flux density is
$
\Pi(S) dS =  | \Pi(\delta) d \delta |$.
\begin{equation}
\Pi(S) = \Pi(\delta) \biggl| \frac {d S } {d \delta} \bigg|^{-1}
\end{equation}
Differentiating Equation~\ref{eqn:observedflux} yields
\begin{equation}
\frac {dS} {d \delta} =  (p - \alpha) \delta^{\,p - \alpha - 1}
\end{equation}
Inserting Equation~\ref{eqn:pofdeltafinal} for isotropically oriented sources
gives
\begin{equation}
\Pi(S) =  [2 \gamma \beta (p -\alpha) \delta^{\,p - \alpha + 1}]^{-1}
\end{equation}
In terms of the observed flux density $S$, 
\begin{equation}
\label{eqn:differentialcount}
\Pi(S) = [2 \gamma \beta (p - \alpha)]^{-1} 
S^{-\bigl(\frac {p - \alpha + 1}{p - \alpha}\bigr)}
\end{equation}
over the flux-density range 
 $[\gamma(1+\beta)]^{\alpha - p} < S <
[\gamma (1+ \beta)^p -{\alpha}$.
Equation~\ref{eqn:differentialcount} specifies the flux-density distribution
of a population of intrinsically identical and
isotropically oriented sources.

For example, consider a population of identical optically selected 
QSOs and assume that the optical emission is unbeamed, so the QSOs have
random orientations.  Assume further that each QSO has one radio component
with intrinsic flux density $S(\nu) = 1$ moving 
with Lorentz factor $\gamma$.  If there are $N_0$ QSOs per steradian, 
the differential count $n(S)$ of radio sources per steradian is given by
$n(S) = N_0 \Pi(S)$ or
\begin{equation}
n(S) = \biggl[ \frac {N_0} {2 \gamma \beta ( p - \alpha)} \biggr]
S^{\bigl( \frac{p -\alpha + 1}{\alpha-p} \bigr)}
\end{equation}
The quantity in square brackets is independent of $S$,
so $n(S) \propto S^{\,\zeta}$  is a power law with slope 
$\zeta$ given by
\begin{equation}
\label{eqn:nsslopes}
\zeta = \biggl( \frac {p - \alpha +1}{\alpha - p} \biggr).
\end{equation}
For any combination of $2 < p < 3$ and $-1 < \alpha < +1$, the
{\it flatest} slope occurs with $\zeta = -5/4$, which occurs for
$p = 3$, $\alpha = -1$; the {\it steepest} slope is $\zeta = -2$, 
when $p = 2$, $\alpha = +1$.

\clearpage

\end{document}